\documentclass[a4paper,fleqn,usenatbib,useAMS]{mnras}

\usepackage{graphicx}
\usepackage{amsmath, amssymb}
\usepackage{subfig}
\usepackage{enumitem}
\usepackage{ulem}
\usepackage{xcolor}
\usepackage{caption}
\usepackage{dcolumn}
\usepackage{arydshln}
\usepackage{soul}

\setstcolor{red}

\defcitealias{Hewett10}{HW10}
\defcitealias{Shen11}{S11}
\defcitealias{Shenetal16}{S16}
\newcolumntype{d}[1]{D{.}{.}{#1}}

\title[Evaluating and Improving Quasar Redshifts]{Evaluating and Improving the Redshifts of $z > 2.2$ Quasars}
\author[Mason et al.]{Michelle Mason$^{1}$\thanks{E-mail: mmason10@uwyo.edu}, Michael S. Brotherton$^{1}$, and Adam Myers$^{1}$ \\
$^{1}$Department of Physics and Astronomy, Dept. 3905, University of Wyoming, 1000 E. University, Laramie, WY 82071, USA}

\begin{document}		

\date{}
\pagerange{\pageref{firstpage}--\pageref{lastpage}} \pubyear{2017}
\maketitle
\label{firstpage}

\begin{abstract}
Quasar redshifts require the best possible precision and accuracy for a number of applications, such as setting the velocity scale for outflows as well as measuring small-scale quasar-quasar clustering.  The most reliable redshift standard in luminous quasars is arguably the narrow \mbox{[\ion{O}{iii}] $\lambda\lambda$4959,5007} emission line doublet in the rest-frame optical. We use previously published [\ion{O}{iii}] redshifts obtained using near-infrared spectra in a sample of 45 high-redshift ($z > 2.2$) quasars to evaluate redshift measurement techniques based on rest-frame ultraviolet spectra. At redshifts above $z = 2.2$ the \mbox{\ion{Mg}{ii} $\lambda$2798} emission line is not available in observed-frame optical spectra, and the most prominent unblended and unabsorbed spectral feature available is usually \mbox{\ion{C}{iv}\,$\lambda$1549}.  Peak and centroid measurements of the \ion{C}{iv} profile are often blueshifted relative to the rest-frame of the quasar, which can significantly bias redshift determinations.  We show that redshift determinations for these high-redshift quasars are significantly correlated with the emission-line properties of  \ion{C}{iv} (i.e., the equivalent width, or EW, and the full width at half maximum, or FWHM) as well as the luminosity, which we take from the Sloan Digital Sky Survey Data Release 7.  We demonstrate that empirical corrections based on multiple regression analyses yield significant improvements in both the precision and accuracy of the redshifts of the most distant quasars and are required to establish consistency with redshifts determined in more local quasars.
\end{abstract}



\section{Introduction}

The first recognized quasar, \mbox{3C 273}, had at the time a dramatically high redshift of 0.158, which first revealed the extreme luminosity of the class \citep{Schmidt63}.  Today there are quasars known to have redshift as great as 7  \citep{Mortlock11}.  Measuring quasar redshifts with high accuracy and precision is challenging, especially when using rest-frame ultraviolet (UV).   

Ideally, host galaxy absorption lines could be used to provide good redshifts; however, in luminous quasars, especially in the ultraviolet, these lines are overpowered by the light from the quasar and therefore much too diluted to be useful.  Furthermore, quasar emission lines are shifted relative to each other, with high-ionization lines often showing large blueshifts \citep{Gaskell82}.  This is especially problematic not just in determining distance, but because accurate redshifts are required to set the velocity zero-point for kinematic studies, which are essential for determining the properties of the outflows believed to play a role in feedback \citep[e.g.,][]{Hopkins10}.

The narrow \mbox{[\ion{O}{iii}] $\lambda$5007} line has been identified as a reliable, albeit not perfect, redshift reference for quasars \citep[e.g.][]{Heckman81, Boroson05}.  In more distant quasars ($z > 0.8$) [\ion{O}{iii}] becomes shifted out of the readily available optical part of the spectrum and into the near-infrared (NIR) region, for which there is a factor of about a thousand times less spectra  available than in the optical.  The most dependable strong UV emission feature that has a peak redshift close to systemic is \mbox{\ion{Mg}{ii} $\lambda$2798} \citep[e.g.,][]{Tytler92}.  For $z > 2.2$ the observed-frame wavelength of \ion{Mg}{ii} moves beyond 9000\,\AA.  This makes it challenging to observe against atmospheric lines, and also pushes it outside of the main sensitivity range of many large spectroscopic surveys.  At these high redshifts there are not many good spectral lines to use as redshift indicators.  The strongest, least-blended, least-absorbed emission feature seen in the observed-frame optical is the \mbox{\ion{C}{iv} $\lambda$1549} doublet.  However, using \ion{C}{iv} presents challenges, and most recent approaches have avoided using \ion{C}{iv} by itself in favor of other techniques.  

The Sloan Digital Sky Survey (SDSS), with a series of data releases, has now cataloged several hundred thousand quasars \citep[e.g.,][]{Schneider10, Paris14, Paris16}.  The SDSS-I/II spectra only have coverage out to about 9000\,\AA, while the SDSS-III/IV go out to about 10000\,\AA.  In this paper we use SDSS-I/II data, specifically from the Seventh Data Release (DR7), which has well tested spectral measurements available \citep{Shen11}.  The SDSS-I/II assigns redshifts using an automated pipeline process of cross-correlation against mean spectra, which gives weight to the entire spectrum rather than just one or a few spectral lines where the majority of the redshift information is contained. Cross-correlation techniques are influenced by blueshifted lines and also produce abrupt redshift discontinuities as different lines shift into and out of the spectral window.  \citet[hereafter \citetalias{Hewett10}]{Hewett10} tackled this problem statistically and applied a number of corrections to the quasar redshifts for SDSS DR7, including a luminosity term. With a sample of several thousand quasars their results are statistically robust, however their approach did not take into account any information about the emission-line profiles, and they acknowledge that their redshifts are less reliable when \ion{Mg}{ii} is not present, as in the case of high-redshift ($z > 2.2$) spectra from SDSS-I/II.

\citet{Brotherton94a} first showed that the blueshifts of the \ion{C}{iv}-emission peak and the peak of the \mbox{\ion{C}{iii}] $\lambda$1909}+\mbox{\ion{Si}{iii} $\lambda$1892} blend relative to \ion{Mg}{ii} appear to be correlated with properties of the line profiles themselves: their peaks are systematically more blueshifted when the lines have larger FWHM and smaller EW \citep[see also][]{Wills93}.  These trends are related to the optical ``Eigenvector 1'' trends \citep[e.g.,][]{Boroson92, Brotherton99, Wills99} and are most likely associated with the accretion rate characterized by the Eddington fraction \citep{Boroson02, Shen14, Sun15} and the modified Baldwin effect \citep{Shemmer15}.  Physically, they appear to result from the systematic variation of a non-reverberating intermediate-line region (ILR) component \citep{Brotherton94b, Denney12} relative to a broader, blueshifted component that varies in velocity width with orientation \citep{Runnoe14}.  The correlations among the FWHM, EW, and blueshift of \ion{C}{iv} are present in SDSS quasar samples \citep[e.g.,][]{Richards11, Coatman17}. 

Here we use measurements based on NIR spectra of high-redshift quasars to not only evaluate several redshift formulas used in practice, but to also develop improvements using \ion{C}{iv} profile measurements to obtain the most accurate redshifts.  We have compiled a small catalog of NIR spectra with [\ion{O}{iii}] measurements, which we use as our baseline.  We use multiple regression analysis with \ion{C}{iv} line profiles in order to predict [\ion{O}{iii}] redshifts.  We also show how other redshift determination methods can be improved by implementing a similar strategy. 

In \S\,\ref{sec:sample} we describe our 45-object sample chosen from the literature, including the original target selection and the cuts we made to the data.  In \S\,\ref{sec:nir_analysis} we discuss the various methods of determining quasar redshifts and how they can be improved using regression analyses to better predict the redshift.  In \S\,\ref{sec:discussion} we discuss our results and justify their importance to quasar astronomy.  Lastly in \S\,\ref{sec:conclusion} we summarize our conclusions.  Throughout this paper, in order to be consistent with \citet[hereafter \citetalias{Shen11}]{Shen11} we use cosmological parameters $\Omega_{\Lambda}$ = 0.7,  $\Omega_{m}$ = 0.3, and $H_{0}$=70 km\,s$^{-1}$ Mpc$^{-1}$.


\section{Sample}
\label{sec:sample}

For our analyses we compiled a sample drawn from the literature \citep{Netzer07, Zuo15, Shen16} that includes high-quality measurements of the rest-frame optical spectra of quasars without \ion{Mg}{ii} in observed-frame optical spectra (generally with $z > 2.2$).  We used these literature references to provide peak [\ion{O}{iii}] redshifts as the most reliable redshift baseline to test UV redshift determinations.  All objects in our sample are included in the SDSS DR7 quasar catalog and have C IV measurements \citep{Shen11}, so we were able to extract \ion{C}{iv} FWHM and EW along with luminosity (log$L_{1350}$) measurements for each object.  

A detailed description of target selection can be found in each literature source.  We will give a brief description here. \citet{Netzer07} observed bright 15 high-redshift, high-luminosity quasars whose spectra included both the H$\beta$ and 5100\,\AA\ continuum features.  To obtain redshift from [\ion{O}{iii}], they assumed that the continuum-subtracted $\lambda 5007$ and $\lambda 4959$ had equal FWHM and an intensity ratio of 3:1.  They then fit both lines with a Gaussian profile and determined the line peak redshift for $\lambda 5007$ by taking the position of the maximum of the fit.  \citet{Zuo15} selected their targets from SDSS DR7 my constraining the redshift ($3.2 < z <3.8$) and magnitude ($m_{\rm{J}} < 17$ and $m_{\rm{K}} < 16$) where both H$\beta$ and \ion{Mg}{ii} were available.  With 32 objects they also fit Gaussians to the [\ion{O}{iii}] doublet and determined the redshift from the line center offset.  \citet{Shen16} determined the shift of the $\lambda 5007$ line peak of 14 $z \sim 3.3$ quasars.  Their process involved fitting the line with two Gaussians: one for the ``core'' and one for the blue ``wing.''  The redshift was determined from the model's peak.  Together, these sources provided 61 data points.  However, 4 objects (SDSSJ025021-075750, SDSSJ025905+001121, SDSSJ030449-000813, and SDSSJ094202+042244) overlapped between the \citet{Zuo15} and \citet{Shen16} samples; we elected to take our data from the \citet{Shen16} sample because they reported greater precision.  There was 1 outlying object (SDSSJ2238-0921) from \citet{Shen16} where an iron feature was misidentified as [\ion{O}{iii}] and provided an unreliable redshift and therefore removed.  This  brought our sample down to 56 objects.  

\begin{figure}
\centering
\includegraphics[width= \columnwidth]{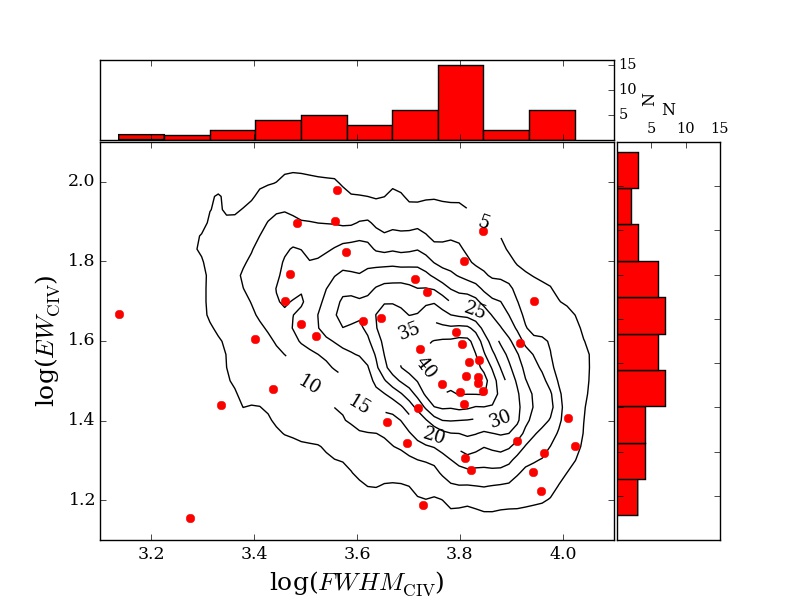}
\caption{Black contours map just over 17,000 $z>2.2$ quasars from the SDSS DR7 catalog and show their distribution in C IV FWHM-EW space.  Red dots indicate the quasars in our NIR sample (which are also in DR7).  The NIR quasars are fairly uniformly spread out and not clustered together, indicating that there is no obvious bias in our selected sample.}
\label{fig:unbiased}
\end{figure}

One object from \citet{Zuo15} was not in SDSS DR7 and was removed from our sample since \ion{C}{iv} measurements were not available.  We eliminated 9 objects with broad absorption lines (BALs) because their measurements of FWHM and EW are not reliable.  We also removed one weak line quasar (WLQ, typically \mbox{EW $<$ 10 \AA}) since WLQs typically behave differently, or are more challenging to measure, than other quasars \citep{Shemmer15}.  Our final NIR sample consisted of 45 quasars.  Figure~\ref{fig:unbiased} shows that these objects span the entire FWHM-EW space occupied by the DR7 catalog, thus indicating that our sample is not strongly biased in any obvious way, although we do have a few outliers.  Table~\ref{tab:nir} shows the data for our sample.

\begin{table*}
\centering
\begin{minipage}{0.95\textwidth}
\centering
\caption{Our compiled NIR sample using objects from literature sources: \citet[denoted N07]{Netzer07}, \citet[denoted Z15]{Zuo15}, and \citet[denoted Sh16]{Shen16}.  All $z_{[\ion{O}{iii}]}$ measurements are taken from literature sources, while $z_{\rm{pipe}}$, $z_{\rm{HW}}$, $FWHM_{\ion{C}{iv}}$, $EW_{\ion{C}{iv}}$, $v_{\ion{C}{iv}}$, and log($L_{1350})$ are taken from SDSS DR7 \citep{Shen11}.}
\label{tab:nir}
\begin{tabular*}{0.95\textwidth}{ l c d{2.4} c c d{5.1} d{3.2} d{4.1} d{2.1}}
\hline \hline
Object ID (SDSS J)   &   Reference   &   \multicolumn{1}{c}{$z_{[\ion{O}{iii}]}$}   &   $z_{\rm{pipe}}$   &   $z_{\rm{HW}}$   &   \multicolumn{1}{c}{$FWHM_{\ion{C}{iv}}$}   &   \multicolumn{1}{c}{$EW_{\ion{C}{iv}}$}   &   \multicolumn{1}{c}{$\Delta v_{\ion{C}{iv}}$}   &   \multicolumn{1}{c}{\rm{log($L_{1350}$)}}  \\
 &  &  &  &  & \multicolumn{1}{c}{(km\,s$^{-1}$)} & \multicolumn{1}{c}{$\rm{(\AA)}$} & \multicolumn{1}{c}{(km\,s$^{-1}$)} & \multicolumn{1}{c}{$\rm{(L_{\odot})}$}  \\
\hline
011521.20+152453.3 & Z15 & 3.443 & 3.4150 & 3.432551 & 6562 & 35.2 & -413 & 46.940 \\
012403.77+004432.6 & Z15 & 3.834 & 3.8343 & 3.827481 & 8248 & 39.4 & -1651 & 47.123 \\
014049.18$-$083942.5 & Z15 & 3.717 & 3.7131 & 3.726035 & 4562 & 25.0 & -413 & 47.263 \\
014214.75+002324.2 & Z15 & 3.379 & 3.3704 & 3.373789 & 5231 & 27.0 & -733 & 46.997 \\
015741.57$-$010629.6 & Z15 & 3.572 & 3.5645 & 3.571022 & 5441 & 52.8 & -637 & 46.998 \\
025021.76$-$075749.9 & Sh16 & 3.3376 & 3.3443 & 3.344000 & 6620 & 18.9 & -691 & 47.041 \\
025438.36+002132.7 & N07 & 2.456 & 2.4629 & 2.463708 & 2879 & 50.2 & 144 & 46.061 \\
025905.63+001121.9 & Sh16 & 3.3724 & 3.3652 & 3.376630 & 3795 & 66.5 & 358 & 47.187 \\
030341.04$-$002321.9 & Z15 & 3.233 & 3.2267 & 3.234527 & 6373 & 39.1 & -637 & 47.234 \\
030449.85$-$000813.4 & Sh16 & 3.2859 & 3.2951 & 3.296364 & 2167 & 27.5 & -723 & 47.379 \\
035220.69$-$051702.6 & Sh16 & 3.2892 & 3.2614 & 3.271370 & 6857 & 31.2 & -819 & 46.699 \\
075303.34+423130.8 & Z15 & 3.590 & 3.5900 & 3.594747 & 2728 & 30.1 & 230 & 47.125 \\
075819.70+202300.9 & Z15 & 3.761 & 3.7471 & 3.753027 & 6493 & 32.6 & -1065 & 47.026 \\
080430.56+542041.1 & Z15 & 3.759 & 3.7592 & 3.755210 & 6885 & 35.6 & -1662 & 47.115 \\
080819.69+373047.3 & Z15 & 3.480 & 3.4717 & 3.476813 & 6313 & 29.7 & -1257 & 46.782 \\
080956.02+502000.9 & Z15 & 3.281 & 3.2813 & 3.287604 & 4094 & 44.7 & -113 & 46.978 \\
081011.97+093648.2 & Sh16 & 3.3906 & 3.3663 & 3.386692 & 10545 & 21.7 & 176 & 46.869 \\
081855.77+095848.0 & Z15 & 3.700 & 3.6736 & 3.688469 & 6993 & 29.9 & -883 & 47.078 \\
082535.19+512706.3 & Z15 & 3.512 & 3.5118 & 3.507203 & 6443 & 20.2 & -1555 & 47.025 \\
083630.54+062044.8 & N07 & 3.397 & 3.4000 & 3.384170 & 5355 & 15.4 & -2045 & 46.294 \\
090033.50+421547.0 & Z15 & 3.290 & 3.2901 & 3.293742 & 4427 & 45.6 & -263 & 47.517 \\
091054.79+023704.5 & Sh16 & 3.2902 & 3.2951 & 3.290458 & 5822 & 31.1 & -1544 & 46.747 \\
094202.04+042244.5 & Sh16 & 3.2790 & 3.2755 & 3.284280 & 3317 & 41.0 & -145 & 47.366 \\
095141.33+013259.5 & N07 & 2.411 & 2.4284 & 2.419197 & 3637 & 95.1 & -1054 & 45.699 \\
095434.93+091519.6 & Sh16 & 3.4076 & 3.3817 & 3.397527 & 10238 & 25.5 & 273 & 46.693 \\
100710.70+042119.2 & N07 & 2.363 & 2.3641 & 2.366708 & 5157 & 57.1 & -1225 & 45.898 \\
101257.52+025933.1 & N07 & 2.434 & 2.4330 & 2.440567 & 6424 & 63.4 & -830 & 46.094 \\
101908.26+025431.9 & Sh16 & 3.3829 & 3.3907 & 3.379110 & 6837 & 32.4 & -1502 & 46.980 \\
103456.31+035859.4 & Sh16 & 3.3918 & 3.3696 & 3.388422 & 6432 & 27.7 & -102 & 47.044 \\
105511.99+020751.9 & N07 & 3.391 & 3.3839 & 3.404447 & 6991 & 75.2 & -242 & 46.374 \\
113838.27$-$020607.2 & N07 & 3.352 & 3.3432 & 3.347426 & 6194 & 41.9 & -562 & 46.386 \\
115111.20+034048.2 & N07 & 2.337 & 2.3370 & 2.336526 & 2942 & 58.6 & -17 & 45.359 \\
115304.62+035951.5 & N07 & 3.426 & 3.4322 & 3.437321 & 1887 & 14.3 & -477 & 46.531 \\
115935.63+042420.0 & N07 & 3.451 & 3.4480 & 3.455751 & 3102 & 44.0 & -27 & 46.599 \\
115954.33+201921.1 & Z15 & 3.426 & 3.4260 & 3.432442 & 8744 & 18.6 & -1640 & 47.394 \\
125034.41$-$010510.6 & N07 & 2.397 & 2.3975 & 2.398533 & 3612 & 79.8 & -402 & 45.895 \\
144245.66$-$024250.1 & N07 & 2.356 & 2.3252 & 2.354779 & 8792 & 50.1 & 1605 & 46.071 \\
153725.35$-$014650.3 & N07 & 3.452 & 3.4522 & 3.467338 & 5279 & 38.0 & -819 & 46.484 \\
173352.23+540030.4 & Z15 & 3.432 & 3.4248 & 3.435445 & 4971 & 22.1 & -327 & 47.418 \\
210258.22+002023.4 & N07 & 3.328 & 3.3448 & 3.341787 & 1369 & 46.5 & -776 & 46.125 \\
213023.61+122252.2 & Z15 & 3.272 & 3.2717 & 3.278919 & 2519 & 40.2 & -92 & 47.098 \\
224956.08+000218.0 & Z15 & 3.311 & 3.3110 & 3.323341 & 3046 & 79.0 & -27 & 46.752 \\
230301.45$-$093930.7 & Z15 & 3.492 & 3.4550 & 3.470379 & 9204 & 20.9 & -349 & 47.250 \\
232735.67$-$091625.6 & Z15 & 3.263 & 3.2626 & 3.281305 & 8161 & 22.3 & 283 & 46.801 \\
234625.66$-$001600.4 & Z15 & 3.507 & 3.4895 & 3.477123 & 9058 & 16.7 & -3086 & 47.140 \\
\hline
\end{tabular*}
\end{minipage}
\end{table*}


\section{Analysis \& Results}
\label{sec:nir_analysis}

\subsection{Methods to Determine UV Redshifts}
We first evaluate several methods proposed or regularly used to determine the redshift from an ultraviolet spectrum that includes \ion{C}{iv} but not \ion{Mg}{ii}.  The most straightforward way to get a redshift would be to simply use the peak of the \ion{C}{iv} emission line, which unfortunately is often known to suffer from large blueshifts.  The SDSS-I/II uses, as part of its pipeline, a cross-correlation technique.  \citetalias{Hewett10} modified the SDSS-I/II cross-correlation technique to correct for luminosity trends.  And finally, \citet[hereafter \citetalias{Shenetal16}]{Shenetal16} provide a correction for the \ion{C}{iv} blueshift based on luminosity that can be applied to the \ion{C}{iv} redshift to obtain an improved redshift.  Before expanding on these methods with our own approach, we make note of some sign conventions and the conversion between redshift and velocity shifts:
\begin{equation}
\frac{\Delta v}{c} = \frac{\Delta z}{1+z}
\label{eqn:dvdz}
\end{equation}
and $z_{\rm{sys}} = z_{\rm{meas}} - \Delta z$, where positive values of $\Delta v$, and therefore $\Delta z$, indicate redshift, while negative values indicate blueshift.

\begin{itemize}[leftmargin=*]
\item \textsc{\ion{C}{iv} line peak} \\
When there are no other prominent spectral lines visible in the observing window it could be necessary to determine the quasar redshift using only the \ion{C}{iv} line peak.  \citetalias{Shen11} reports velocity shifts of several spectral lines relative to the systemic redshifts (which are pipeline redshifts, as we will discuss next) reported in SDSS DR7 by \citet{Schneider10}.  These velocity shifts are determined from the peak of a multiple-Gaussian model.  For evaluation purposes we use the \ion{C}{iv} velocity shift and Equation\,\ref{eqn:dvdz}, in conjunction with the SDSS-I/II pipeline redshift, to recover the \ion{C}{iv} redshift.  

\item \textsc{SDSS-I/II Pipeline Cross-Correlation} \\
The SDSS-I/II pipeline uses a cross-correlation method, following the \citet{Tonry79} technique and using the composite quasar templates from \citet{VandenBerk01}.  Each quasar spectrum is continuum-subtracted before being Fourier-transformed and then compared to the transform of each template spectra.  The cross-correlation redshift is determined by which cross-correlation function has the highest confidence limit amid all template comparisons.  While most quasars exhibit a rather featureless continuum with a few broad lines in the optical through UV, there are still spectral variations that this method of template-fitting does not always match, and which affect the resulting redshift value.  \citet{Schneider10} reports the pipeline redshifts for DR7 quasars.  

\item \textsc{\citet{Hewett10} Redshifts} \\
\citetalias{Hewett10} devised a new cross-correlation method in order to improve on the SDSS-I/II pipeline values by correcting the zero-point over a large range in redshift.  This new method includes both systematic and luminosity-dependent corrections to account for the slight blueshift of [\ion{O}{iii}] as well as the relative shift between emission lines.  \citetalias{Hewett10} claims that the so-called Princeton algorithm is more accurate for $z > 2.0$ (about where \ion{Mg}{ii} becomes shifted out of the SDSS-I/II spectral window), but these redshifts are not generally readily available, and not verified against [\ion{O}{iii}] or other more reliable redshift indicators.

\item \textsc{\citet{Shenetal16} Velocity Shift Correlations} \\
\citetalias{Shenetal16} looks at the velocity shifts between multiple pairs of line peaks and presents a simple method to correct the mean redshift offset as a function of quasar luminosity.  For each spectrum they subtract the continuum and then fit multiple-Gaussian models to the resulting line spectrum.  Their results indicate that the average shift between lines goes as $\Delta v = a + b\left( \mathrm{log}L - \mathrm{log}L_0 \right)$ where $a$ is the mean velocity shift at the reference luminosity (log$L_0$), $b$ is the luminosity dependence, and log$L$ is the specific monochromatic continuum luminosity for the emission-line pair.  Statistical tests show that the luminosity correlation is only significant for \ion{He}{ii}, \ion{Si}{iv}, and \ion{C}{iv}.  To determine the velocity offset of \ion{C}{iv} relative to [\ion{O}{iii}] following this technique, one must employ a bootstrapping method and step through several line pairs provided:
\begin{equation}
\begin{split}
\Delta v_{\ion{C}{iv} - [\ion{O}{iii}]} = \: & \Delta v_{ \ion{C}{iv} - \ion{Mg}{ii} } + \Delta v_{ \ion{Mg}{ii} - [\ion{O}{ii}] } +   \\ 
& \Delta v_{ [\ion{O}{ii}] - \ion{Ca}{ii} } - \Delta v_{ [\ion{O}{iii}] - \ion{Ca}{ii} } \,.
\end{split}
\end{equation}
Here, only $\Delta v_{\ion{C}{iv} - \ion{Mg}{ii}}$ requires the luminosity term $b$ because it is the only line pair that has a significant luminosity effect.  Once the final $\Delta v$ is calculated it can be evaluated directly by comparison to the measured shift of \ion{C}{iv} relative to [\ion{O}{iii}].
\end{itemize}

\begin{figure}
\centering
\includegraphics[trim={5mm 2cm 5mm 0}, clip, width=\columnwidth]{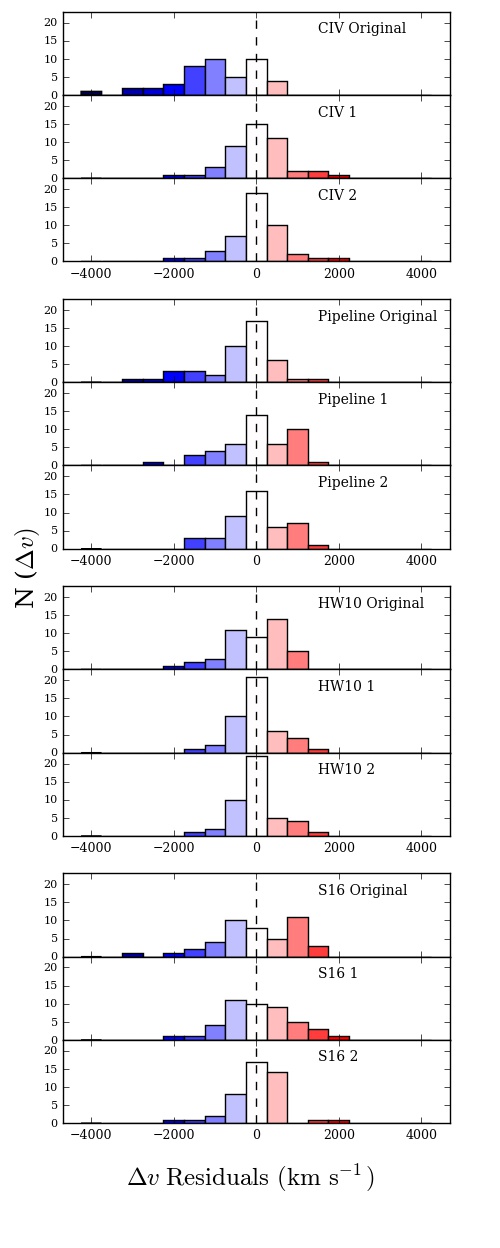}
\caption{$\Delta v_{ \ion{C}{iv} - [\ion{O}{iii}] }$ for each redshift prediction method are displayed.  Unbiased predictions should be symmetric about the dashed vertical line ($\Delta v =0$), and more reliable predictions have a narrower distribution.  Table\,\ref{tab:stats} shows statistics for each histogram.}
\label{fig:hist}
\end{figure}

We have calculated the velocity offset $\Delta v_{ \ion{C}{iv} - [\ion{O}{iii}] }$ using each of these four methods and refer to them as the ``Original'' predictions.  The redshifts reported by \citetalias{Shen11} and \citetalias{Hewett10} are not $z_{[\ion{O}{iii}]}$ but rather $z_{\rm{systemic}}$; to recover $z_{[\ion{O}{iii}]}$ we incorporate a recommended blueshift of 45 km\,s$^{-1}$ \citep{Hewett10}.  The mean, median, spread, and skew of each Original prediction were calculated using basic Python packages (\texttt{numpy} and \texttt{scipy}) and are shown in Table\,\ref{tab:stats}.  The distribution of velocity offsets are shown in the top panels in Figure\,\ref{fig:hist}.  

\begin{table}
\centering
\caption{Statistics for $\Delta v_{ \ion{C}{iv} - [\ion{O}{iii}]}$ for each redshift prediction.  It is clearly shown that Corrections 1 and 2 drastically improve \ion{C}{iv} velocity offset predictions over the current methods that are commonly used.  For reference, a perfect prediction would have a mean, median, spread, and skew all equal to zero.}
\label{tab:stats}
\begin{tabular}{l | d{3.1}  d{2.0}  d{3.0} d{1.2} }
\hline \hline
Model   &   \multicolumn{1}{c}{Mean}   &   \multicolumn{1}{c}{Median}   &   \multicolumn{1}{c}{Spread}   &   \multicolumn{1}{c}{Skew} \\
    &   \multicolumn{1}{c}{\footnotesize{(km\,s$^{-1}$)}}   &   \multicolumn{1}{c}{\footnotesize{(km\,s$^{-1}$)}}   &   \multicolumn{1}{c}{\footnotesize{(km\,s$^{-1}$)}}   &   \\
\hline
\ion{C}{iv} Original & -985 & -971 & 999 & -0.77 \\ 
\ion{C}{iv} 1 & 16 & 67 & 762 & -0.07 \\ 
\ion{C}{iv} 2 & 1 & 23 & 673 & 0.06 \\ [1ex]
Pipeline Original & -414 & -73 & 891 & -0.75 \\ 
Pipeline 1 & 17 & 8 & 809 & -0.70 \\ 
Pipeline 2 & 1 & 11 & 716 & -0.17 \\ [1ex]
HW Original & 6 & 90 & 672 & -0.74 \\ 
HW 1 & 9 & 8 & 577 & 0.22 \\ 
HW 2 & 0 & -52 & 541 & 0.45 \\ [1ex]
Shen Original & 14 & 91 & 962 & -0.74 \\ 
Shen 1 & 19 & 47 & 815 & -0.11 \\ 
Shen 2 & -1 & -44 & 675 & -0.06 \\ [1ex]
\hline
\end{tabular}
\end{table}

\subsection{Correlations with FWHM \& EW}
\begin{figure*}
\centering
\begin{minipage}{\textwidth}
\centering
\includegraphics[height=0.95\textheight]{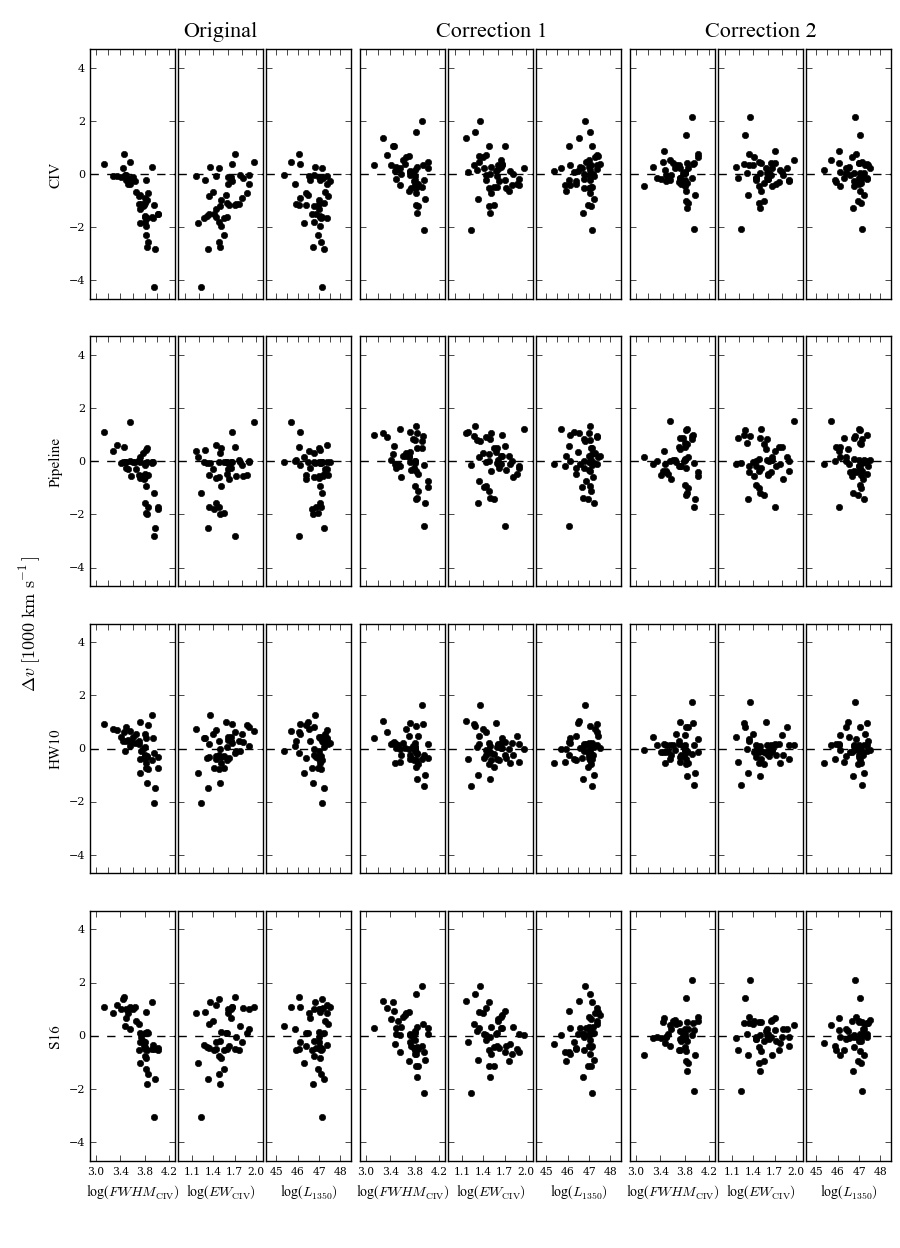}
\vspace{-7mm}
\caption{Our regression parameters, \ion{C}{iv} FWHM, \ion{C}{iv} EW, and luminosity, plotted against $\Delta v_{\ion{C}{iv} - [\ion{O}{iii}]}$ for several redshift determinations and modified versions using Equations\,\ref{eqn:corr1} and \ref{eqn:corr2} using the coefficients of Table\,\ref{tab:coeff}.  From top to bottom, we plot the velocity differences between [\ion{O}{iii}] and UV redshifts determined from the peak of \ion{C}{iv}, the SDSS-I/II pipeline, \citetalias{Hewett10}, and \citetalias{Shenetal16}.  The left-most ``Original'' columns refer to velocity shifts that have not had our regression routine applied, ``Correction 1'' are velocity shifts that have been corrected for FWHM and EW of \ion{C}{iv}, and ``Correction 2'' refers to velocity shifts that have also been corrected for luminosity.  The "Original" columns show some very significant collations, which are largely eliminated in the corrected columns.}

\label{fig:trends}
\end{minipage}
\end{figure*}

Upon inspection, each Original velocity offset shows a trend in velocity offset relative to [\ion{O}{iii}] with respect to both FWHM and EW of \ion{C}{iv}.  These trends are displayed in the left-most columns of Figure\,\ref{fig:trends}.  Correlation coefficients and significances for these trends are shown in Table\,\ref{tab:correlations}.  Predictions that incorporated luminosity show a lower correlation coefficient for EW since luminosity correlates with EW through the Baldwin effect and is therefore already slightly accounted for \citep{Baldwin77}.   

\begin{table*}
\centering
\begin{minipage}{0.75\textwidth}
\centering
\caption{Spearman's rank correlation coefficients (R) and significances (P).  As shown, the values of the correlation coefficients typically become smaller in magnitude when our regression routine is applied, indicating that we are successfully removing the effect of these parameters on the velocity residuals.}
\label{tab:correlations}
\begin{tabular*}{0.87\textwidth}{ l | d{3.2} d{3.6} d{3.2} d{3.6}  d{3.2} d{3.6} }
\hline \hline
Model   &   \multicolumn{1}{c}{$\rm{R_{FWHM}}$}   &  \multicolumn{1}{c}{$\rm{P_{FWHM}}$}   &   \multicolumn{1}{c}{$\rm{R_{EW}}$}   &  \multicolumn{1}{c}{$\rm{P_{EW}}$} &   \multicolumn{1}{c}{$\rm{R_{L}}$}   &  \multicolumn{1}{c}{$\rm{P_{L}}$}\\
\hline
\ion{C}{iv} Original  &  -0.74  &  7.86E-09  &  0.44  &  2.71E-03  &  -0.23  &  0.13 \\ 
\ion{C}{iv} 1  &  -0.43  &  3.32E-03  &  -0.21  &  0.17  &  0.11  &  0.48 \\ 
\ion{C}{iv} 2  &  -0.01  &  0.94  &  -0.07  &  0.63  &  -0.11  &  0.47 \\ [1ex]
Pipeline Original  &  -0.54  &  1.11E-04  &  0.17  &  0.27  &  -0.20  &  0.18 \\ 
Pipeline 1  &  -0.34  &  0.02  &  -0.21  &  0.18  &  -0.00  &  0.98 \\ 
Pipeline 1  &  -0.01  &  0.97  &  -0.01  &  0.96  &  -0.18  &  0.23 \\ [1ex]
HW10 Original  &  -0.57  &  4.64E-05  &  0.34  &  0.02  &  -0.16  &  0.28 \\ 
HW10 1  &  -0.33  &  0.03  &  -0.11  &  0.47  &  0.09  &  0.57 \\ 
HW10 2  &  -0.07  &  0.66  &  -0.03  &  0.85  &  -0.07  &  0.66 \\ [1ex]
S16 Original  &  -0.71  &  4.59E-08  &  0.32  &  0.03  &  0.03  &  0.85 \\ 
S16 1  &  -0.45  &  1.72E-03  &  -0.21  &  0.16  &  0.31  &  0.04 \\ 
S16 2  &  0.01  &  0.97  &  -0.02  &  0.89  &  0.16  &  0.30 \\ [1ex]
\hline
\end{tabular*}
\end{minipage}
\end{table*}

\subsection{Multiple Regression}
\label{subsec:regress}
It has long been known that the \ion{C}{iv} peak is blueshifted relative to the systemic redshift of a quasar and that this blueshift correlates with the FWHM and EW of \ion{C}{iv}.  To correct for this, we performed a straightforward multiple regression analysis (Python's \texttt{statsmodels.api.OLS}) on our NIR sample in order to accurately predict the blueshift of the \ion{C}{iv} line using its FWHM and EW as the predictors.  The formula to predict $\Delta v_{\ion{C}{iv}}$ is 
\begin{equation}
\Delta v_{\ion{C}{iv}} = \alpha \mathrm{log}_{10}\left(FWHM_{\ion{C}{iv}}\right) +  \beta \mathrm{log}_{10}\left(EW_{\ion{C}{iv}}\right) .  
\label{eqn:corr1}
\end{equation}
In order to calculate the true redshift of a quasar given only the \ion{C}{iv} line we use $z_{\rm{actual}} = z_{\ion{C}{iv}, \rm{meas}} - \Delta z_{\ion{C}{iv}}$ where $\Delta z$ is given by Equation~\ref{eqn:dvdz} and the coefficients to calculate $\Delta v_{\ion{C}{iv}}$ are given in Table~\ref{tab:coeff}.  We refer to our multiple regression analysis using only FWHM and EW of \ion{C}{iv} as ``Correction 1.''  We also performed our analysis incorporating luminosity (log$L_{1350}$) as a factor in addition to FWHM and EW, which we call ``Correction 2.''  To predict $\Delta v_{\ion{C}{iv}}$ using luminosity use  
\begin{equation}
\begin{split}
\Delta v_{\ion{C}{iv}} = \: & \alpha \mathrm{log}_{10}\left(FWHM_{\ion{C}{iv}}\right) +  \beta \mathrm{log}_{10}\left(EW_{\ion{C}{iv}}\right) + \\ 
& \gamma \mathrm{log}_{10}\left(L_{1350}\right) .  
\end{split}
\label{eqn:corr2}
\end{equation}
The differences in velocity offsets given by Corrections 1 and 2 are not generally large, indicating that the luminosity effect can be mostly accounted for using only FWHM and EW, if necessary.  

Since each of the three previously-mentioned redshift prediction methods show trends with FWHM and EW, and to some extent luminosity, we performed this multiple regression analysis on their corrected redshift values in order to eliminate these trends.  Figure\,\ref{fig:hist} shows the improved velocity offset distributions and Table\,\ref{tab:stats} shows the statistics for these corrections.  Our multiple regression routine improves the straight \ion{C}{iv} velocity shift by $\sim$1000 km\,s$^{-1}$ for both Corrections 1 and 2, and the SDSS-I/II pipeline can be improved by over 400 km\,s$^{-1}$ using this method.  The \citetalias{Hewett10} and \citetalias{Shenetal16} methods already do an excellent job at eliminating the average velocity offset.  While their statistics are slightly improved with this method, the more important result is the removal of FWHM and EW trends, as shown in Figure\,\ref{fig:trends}.  Table\,\ref{tab:correlations} shows the reduced correlation coefficients for these parameters.  It is clear that using a luminosity correction alone is not sufficient and that more information about the line profile itself can be used in order to determine a better velocity offset.  

We acknowledge that this is a small, incomplete sample and that further work done with a larger sample may yield even better results.  For instance, \citetalias{Hewett10} determined a luminosity correction based on the very large and largely complete SDSS quasar sample.  Table\,\ref{tab:stats} shows that the \citetalias{Hewett10} Correction 2 to be slightly better in terms of the spread than Correction 1, but that is primarily an artifact of having an extra free parameter and is not very significant.  Moreover, small differences in the luminosity correction using our much smaller sample compared to \citetalias{Hewett10} are not unexpected statistically.  We therefore recommend using only Correction 1 in the case of \citetalias{Hewett10} redshifts.

We ﬁnd that implementing a luminosity correction for \citetalias{Shenetal16} does more significantly improve the velocity offset prediction, despite the fact that they  already include a luminosity factor. Table\,\ref{tab:stats} shows a significant reduction in the uncertainty for Correction 2 as compared to Correction 1. Furthermore, Table\,\ref{tab:coeff} shows a very significant luminosity term, however it is at the expense of the EW term, likely due to a degree of degeneracy of these parameters resulting from the the Baldwin effect. \citetalias{Shenetal16} had a sample of $\sim$500 objects, and while this sample is larger than our own, empirically Correction 2 including a luminosity term appears preferred.

\begin{table}
\centering
\caption{Coefficients, uncertainties, and significances for multiple regression predictors.  $FWHM$ ($\alpha$) and $EW$ ($\beta$) are used in Correction 1, while $L_{1350}$ ($\gamma$) is also included in Correction 2.  We recommend using these values in Equations\,\ref{eqn:corr1} and \ref{eqn:corr2} in addition to the listed redshift-correction methods in order to obtain the most accurate and precise velocity shift estimates.}
\label{tab:coeff}
\begin{tabular}{l c d{4.2} d{4.2} d{4.2} }
\hline \hline
Correction  & Coeff &  \multicolumn{1}{c}{Value} & \multicolumn{1}{c}{Error} & \multicolumn{1}{c}{t-value} \\
\hline
\ion{C}{iv} 1 &  &  &  & \\
 & $\alpha$ & -1292 & 204 & -6.3 \\
 & $\beta$ & 2451 & 486 & 5.0 \\
\ion{C}{iv} 2 &  &  &  & \\
 & $\alpha$ & -3031 & 537 & -5.6 \\
 & $\beta$ & 1541 & 508 & 3.0 \\
 & $\gamma$ & 168 & 48 & 3.4 \\
Pipeline 1 &  &  &  & \\
 & $\alpha$ & -705 & 216 & -3.3 \\
 & $\beta$ & 1412 & 516 & 2.7 \\
Pipeline 2 &  &  &  & \\
 & $\alpha$ & -2539 & 571 & -4.4 \\
 & $\beta$ & 452 & 541 & 0.8 \\
 & $\gamma$ & 177 & 52 & 3.4 \\
HW10 1 &  &  &  & \\
 & $\alpha$ & -600 & 154 & -3.9 \\
 & $\beta$ & 1439 & 368 & 3.9 \\
HW10 2 &  &  &  & \\
 & $\alpha$ & -1576 & 431 & -3.7 \\
 & $\beta$ & 928 & 408 & 2.3 \\
 & $\gamma$ & 94 & 39 & 2.4 \\
S16 1 &  &  &  & \\
 & $\alpha$ & -893 & 218 & -4.1 \\
 & $\beta$ & 2140 & 520 & 4.1 \\
S16 2 &  &  &  & \\
 & $\alpha$ & -3121 & 538 & -5.8 \\
 & $\beta$ & 973 & 509 & 1.9 \\
 & $\gamma$ & 215 & 49 & 4.4 \\
\hline
\end{tabular}
\end{table}

\section{Discussion}
\label{sec:discussion}

\subsection{Origin of C IV Blueshifts}
This paper is an empirical approach to obtain the best redshift estimates; at this point we do not offer a physical explanation for the \ion{C}{iv} profile or blueshift.  The \ion{C}{iv} profile generally shows a blue asymmetry even when the line peak is consistent with \ion{Mg}{ii} or [\ion{O}{iii}] redshifts \citep[e.g.,][]{Wills93, Brotherton94a}.  The variation of the contribution of the low-velocity gas, that is, the line core called the ILR component, is the primary driver that determines how much of an effect the blueshift of the higher velocity gas has on the line peak velocity shift.  Other mechanisms create and determine the magnitude of the blueshift of the high-velocity gas.

There are a few existing candidates for the origin of the blueshift in general.  What is required, fundamentally, is some gas with radial motions plus a source of opacity.  Outflows can produce blueshifts, for instance, if the gas on the far side of the system is obscured by a central structure.  Inflows can produce blueshifts as well, for instance, if the infalling material is optically thick and the line emission arises preferentially from the side facing the central ionizing continuum.

Disk-wind models \citep[e.g.,][]{Murray95} as the origin of the broad line region (BLR) in general have become popular as they solve a number of long-standing problems (e.g., the smoothness of broad lines).  They also provide a natural explanation for the commonality of blueshifts as material in the disk midplane may be expected to obscure the far side \citep[e.g.,][]{Richards11}.  A competing idea comes from \citet{Gaskell09}, who proposes that scattering off of infalling material produces the \ion{C}{iv} blueshift.  In either case, the high-velocity gas must also behave like a disk, in the sense that velocities are highest in more edge-on systems as seen for \ion{C}{iv} in radio-loud quasars \citep{Vestergaard00, Runnoe14}.

The exact mechanism causing the \ion{C}{iv} blueshift may be currently unresolved, however we are still able to empirically provide a better way to obtain the magnitude of the blueshift.  All of the commonly-used prescriptions for determining redshifts of distant ($z > 2.2$) quasars can be improved in precision and accuracy by exploiting known biases with the \ion{C}{iv} profile; luminosity corrections alone are found wanting.   By adding the predictors \ion{C}{iv} FWHM, \ion{C}{iv} EW, and to a lesser extent luminosity, our simple multiple regression approach allows for better determination of the velocity offsets relative to [\ion{O}{iii}] (by up to several hundred km\,s$^{-1}$) for the three aforementioned methods. If lines like [\ion{O}{iii}], H$\beta$, or \ion{Mg}{ii} are not available, one of our formulations should be used in conjunction with one of the common UV redshift methods.

\subsection{Primary Applications Requiring Better Redshifts}
There exist a few specific needs for better UV redshifts.  At $z = 2.5$, for example, a shift of 500 km\,s$^{-1}$ corresponds to a comoving distance of $\sim 5 h^{-1}$ Mpc \citep{FontRibera13, Prochaska13}, large enough to be problematic for measurements of small-scale quasar clustering and the transverse proximity effect that require precise conversions from redshift to distance.  As the scale length of the clustering of quasars at $z = 2.5$ is on order $7 h^{-1}$ Mpc and the imprecision of quasar redshifts is similar, precise measurements of quasar clustering at $z > 2.5$ cannot be done in the redshift-space direction \citep[e.g.,][]{White12}. When measuring the clustering of high-redshift quasars, data is usually projected across the line of sight to ameliorate this effect.  This discards information and inhibits critical cosmological constraints such as those based on redshift-space distortions and the pairwise velocity distribution \citep[e.g.,][]{Wang14}.

There is also great interest in determining the properties of quasar outflows identified through UV absorption lines, especially given models for AGN feedback affecting star formation in host galaxies \citep[i.e.,][]{Higginbottom13}.  Accurate redshifts are necessary to determine the velocity of the outflowing material, without which accurate momentum and energy transfer cannot be calculated.  Even if not perfect, the best possible redshifts are needed to analyze individual systems.  These are the ones we provide if only a far-UV spectrum including \ion{C}{iv} is available.

\subsection{Future Work}
We acknowledge that the $z > 2.2$ [\ion{O}{iii}] sample we have employed is heterogeneous and includes some outliers.  A larger, more representative sample with more uniform measurements would likely yield some additional quantitative improvement, and we recommend more near-infrared spectroscopy of high-redshift quasars for this purpose.  A parallel approach, with its own pros and cons, would be to perform a similar analysis of the \ion{C}{iv} shift relative to \ion{Mg}{ii}, which is available in SDSS catalogs in large numbers \citep[e.g.,][]{Shen11}, but with the additional issue of applying an average shift between the \ion{Mg}{ii} and [\ion{O}{iii}] redshifts, which may have systematics not yet fully understood.

\section{Conclusion}
\label{sec:conclusion}
We have analyzed the precision and accuracy of three common redshift prediction methods for high-redshift quasars and found significant room for improvement for quasars at $z > 2.2$.   We show that implementing a simple multiple regression routine using \ion{C}{iv} FWHM, \ion{C}{iv} EW, and sometimes luminosity can produce more reliable predictions of the \ion{C}{iv} blueshift.  To obtain the most accurate prediction of the $\Delta v_{ \ion{C}{iv} - [\ion{O}{iii}] }$ blueshift, we recommend using \textit{Correction 2} (using FWHM, EW, and luminosity) when dealing with the \ion{C}{iv} profile,  SDSS DR7 pipeline redshifts, and \citetalias{Shenetal16} velocity shifts.  When using \citetalias{Hewett10} redshifts we recommend \textit{Correction 1} (only FWHM and EW) because the luminosity factor has already been sufficiently removed.  The equations below provide numerical estimates.  Recall that to correct [\ion{O}{iii}] redshifts to systemic requires an additional 45 km\,s$^{-1}$ blueshift.

\begin{itemize}

\item \, Using \ion{C}{iv} alone:
\begin{equation}
\begin{split}
\Delta v_{\ion{C}{iv}} = \: & -3031\, \mathrm{log}_{10}\left(FWHM_{\ion{C}{iv}}\right) \\  
& +  1541\, \mathrm{log}_{10}\left(EW_{\ion{C}{iv}}\right) + 168\, \mathrm{log}_{10}\left(L_{1350}\right) 
\end{split}
\end{equation}

\item \, Using SDSS DR7 Pipeline redshifts: 
\begin{equation}
\begin{split}
\Delta v_{\ion{C}{iv}} = \: & -2539\, \mathrm{log}_{10}\left(FWHM_{\ion{C}{iv}}\right) \\ 
& + 452\, \mathrm{log}_{10}\left(EW_{\ion{C}{iv}}\right) + 177\, \mathrm{log}_{10}\left(L_{1350}\right) 
\end{split}
\end{equation}

\item \, Using \citetalias{Hewett10} corrected redshifts:
\begin{equation}
\begin{split}
\Delta v_{\ion{C}{iv}} = \: & -600\, \mathrm{log}_{10}\left(FWHM_{\ion{C}{iv}}\right) \\
& + 1439\, \mathrm{log}_{10}\left(EW_{\ion{C}{iv}}\right)
\end{split}
\end{equation}

\item \, Using \citetalias{Shenetal16} predicted velocity offsets:
\begin{equation}
\begin{split}
\Delta v_{\ion{C}{iv}} = \: & -893\, \mathrm{log}_{10}\left(FWHM_{\ion{C}{iv}}\right) \\
& + 2140\, \mathrm{log}_{10}\left(EW_{\ion{C}{iv}}\right)  + 215\, \mathrm{log}_{10}\left(L_{1350}\right) 
\end{split}
\end{equation}

\end{itemize}

We not only are able to improve this prediction by up to several hundred km\,s$^{-1}$, but we also manage to remove trends of this blueshift with respect to both FWHM and EW of \ion{C}{iv} in all four redshift prescriptions.  Having the most dependable way to predict the redshift of high-redshift quasars using traditionally problematic spectral lines can lead to better understandings of both quasar dynamics and small-scale quasar clustering.

\section*{Acknowledgments}
We thank the Wyoming NASA Space Grant Consortium who funded this work through the NASA Grant \#NNX15AI08H.  ADM was partially supported by NSF grant 1515404.


\bibliographystyle{mnras}
\setcitestyle{authoryear,open={((},close={))}}
\bibliography{fix_z}


\label{lastpage}
\end{document}